\documentstyle[12pt]{article}

\begin{document}
\sloppy
\begin{flushright}{UT-766\\ January '96}\end{flushright}

\vskip 1.5 truecm

\centerline{\large{\bf A natural solution to the $\mu$-problem  }}
\centerline{\large{\bf in dynamical supergravity model}}
\vskip .75 truecm
\centerline{\bf Tomohiro Matsuda
\footnote{matsuda@th.phys.titech.ac.jp}}
\vskip .4 truecm
\centerline {\it Department of Physics, University of Tokyo}
\centerline {\it Bunkyo-ku, Tokyo 113,Japan}
\vskip 1. truecm
\makeatletter
\@addtoreset{equation}{section}
\def\theequation{\thesection.\arabic{equation}}
\makeatother
\vskip 1. truecm
\begin{abstract}
\hspace*{\parindent}
The Higgs mixing term coefficient $\mu$ is calculated in the
supersymmetric theory which possesses a non-anomalous $U(1)_{R}$ symmetry
in the limit of global supersymmetry.
In this model, supersymmetry is assumed to be broken by gaugino condensation
in the hidden sector when the supergravity effects are  turned on.
The soft breaking terms in the visible sector
and the $\mu$ term of order the weak scale are produced in a simple
manner.
\end{abstract}
\newpage
\section{Introduction}
\hspace*{\parindent}
The Standard Model(SM) of particle interactions enjoys overwhelming
phenomenological successes, but it does not account for the
gravitational interactions nor does it explain the origin or
naturalness of the electroweak scale $M_{W}\ll M_{pl}$.
These theoretical problems result in a brief that the high-energy
physics should be described by  supergravity.
The hierarchy of mass scales can be naturally explained
if supersymmetry is exact at high energies but becomes spontaneously
broken, above $M_{W}$, by a non-perturbative mechanism.
At low energies, this mechanism should decouple from the observable
physics and supersymmetry would appear to be broken by
explicit soft terms in the effective low-energy Lagrangian.
>From the low-energy point of view, these soft terms - which include the
masses of the super-partners of all known particles - are simply
independent input parameters, just like the gauge and the Yukawa
couplings of the
Standard Model, but from the high-energy point of view, they
are calculable in terms of the supergravity couplings.
Because of its non-renormalizability, supergravity itself
has to be thought of as an effective theory, valid below
the Planck scale $M_{pl}$.
Currently, the best candidate for a consistent theory
governing the physics of energies near the Planck scale
is the superstring theory.
However, in general, we  have no reliable stringy mechanisms that
lead to non-perturbative spontaneous breaking of supersymmetry.
Instead, one generally assumes that the dominant non-perturbative
effects emerges at energies well below $M_{pl}$.
Gaugino condensation in an asymptotically free hidden sector
of the effective supergravity is a prime example of this type
of mechanisms.
These models, however, face another problem of naturalness which
we call the $\mu$-problem where $\mu$ is the coefficient of the
$H_{1}H_{2}$ term in the low energy superpotential and
$H_{1}$ and $H_{2}$ denote the usual Higgs SU(2) doublet chiral
superfields.
In the minimal supersymmetric standard model(MSSM) the matter
content consists of three generations of quarks and lepton superfields
plus two higgs doublets $H_{1}$ and $H_{2}$ of opposite hypercharge.
The most general effective observable
superpotential has the form:
\begin{eqnarray}
  \label{super_MSSM}
  W^{MSSM}&=&W^{0}+W^{\mu-term}\nonumber\\
  &&\left\{
  \begin{array}{ccl}
    W^{0}&=&\sum_{generations}(h_{u}Q_{L}H_{2}u_{R}+h_{d}Q_{L}H_{1}d_{R}
    +h_{e}L_{L}H_{1}e_{R})\\
    W^{\mu-term}&=&\mu H_{1}H_{2}
  \end{array}
  \right.
\end{eqnarray}
where $h_{i}$ presents dimensionless Yukawa coupling constants.
This includes the usual Yukawa couplings plus a possible supersymmetric
mass term for the Higgses.
In addition to these supersymmetric terms, we should include
soft breaking terms of order 1Tev.
For this model to work well, it is known that $\mu$ should not be
too large and its value is estimated to be the same order as the
soft breaking terms.
Once it is accepted that the presence of the $\mu$-term is essential,
immediately a question arises.
Is there any dynamical reason why $\mu$ should be so small
of the order of the electroweak scale?
We should note that, to this respect, the $\mu$-term is different
from the supersymmetry breaking terms so its origin should be
different from the supersymmetry breaking mechanism.
In principle the natural scale of $\mu$ would be $M_{pl}$, but
this would re-introduce the hierarchy problem since the Higgs
scalars get a contribution $\mu^{2}$ to their squared mass.
Thus, any complete explanation of the electroweak breaking scale
must justify the origin of $\mu$.
This is the so-called $\mu$-problem and this has been considered by
several authors\cite{review}.
In this letter we suggest a natural solution to the $\mu$-problem
without introducing complicated non-renormalizable terms
and any additional mechanisms other than  dynamical supersymmetry
breaking in the hidden sector.

\section{A natural solution to the $\mu$-problem}
\hspace*{\parindent}
The model of supersymmetry which we study is based on the continuous
$U(1)_{R}$ symmetry that is extended from  R-parity.
We define this R-symmetry by giving the coordinate of superspace
$\theta$ charge +1/2, all matter fields charge +1/2, and all Higgs
superfields charge 0.
Expansions of the superfields in terms of the component fields then show
that all ordinary particles are R-neutral while all superpartners
carry non-zero R-charge.
For gauginos in the hidden sector, there is an R-symmetry which
is spontaneously broken if a gaugino condensate forms and
leads to a Goldstone mode.
(Here we should note that this R-symmetry should be explicitly
broken when we fine-tune the cosmological constant by adding
a constant to the superpotential and
turn on the supergravity effects.
See also \ref{banks} for more detailed discussions on this point.)
In this case the auxiliary field $\phi$ describing this would-be
mode must be embedded in a chiral superfield $\Phi$ which
is coupled in a supersymmetric way.
(Here we can also consider the auxiliary field $\Phi$ as the compensator
superfield for the anomaly of the R-symmetry.)
In this respect,  we  consider an auxiliary field $\Phi$ which also
couples to the ordinary components of MSSM.
This field $\Phi$ has R-symmetry +1, and its scalar component is
an order parameter of gaugino condensation.
(The construction of the Lagrangian is motivated by ref.\cite{ross})
The general supersymmetric lagrangian is generally characterized by
three functions; K\"ahler potential $K(z_{i},z_{i}^*)$, superpotential
$W(z_{i})$, and kinetic function $f(z_{i})$ for vector multiplets.
K\"ahler potential $K(z_{i},z_{i}^*)$ is a function of scalar
fields $z_{i}$ and $z_{i}^*$, while  superpotential $W(z_{i})$ and the
kinetic function $f(z_{i})$ depend scalar fields with definite chirality.
Using these functions, the general form of the supersymmetry
lagrangian can be written  in the  superfield formalism,
\begin{eqnarray}
  \label{lagsugra1}
  L&=&\int d^{4}\theta  K \nonumber\\
  &&+\int d^{2}\theta\left[W+\left\{\frac{1}{4}f_{HS}{\cal WW}
   \right\}_{H.S.}+\left\{\frac{1}{4}f_{VS}{\cal WW}
   \right\}_{V.S.}\right]
    +h.c.
\end{eqnarray}
where $f_{HS}$ and $f_{VS}$ denote the kinetic functions
for the hidden and the visible sector.
Here we simply assume that the hidden sector consists of  $SU(N_{c})$
supersymmetric pure Yang-Mills.
Demanding that the effective theory has non-anomalous $U(1)_{R}$ invariance
in the limit of global supersymmetry, which is compensated by  the auxiliary
field $\Phi$, the form of the
$W$ and $f_{HS}$ are determined:
\begin{eqnarray}
  W&=&W(\Phi)+W^{0}\nonumber\\
  &&W(\Phi)=(m^{2}+\lambda H_{1}H_{2})\Phi\nonumber\\
  f_{HS}&=&  S - \xi ln(\Phi/\mu)
\end{eqnarray}
where $m$ and $\mu$ are the mass parameters, $\xi$ is a dimensionless
constant and  $S$ is a dilaton  superfield.
$W$ is the total superpotential which can be written in the sum of
the hidden, observable and mixing terms.
One can determine $\xi$  by demanding that
the low-energy effective Lagrangian is anomaly free under the
non-anomalous R-symmetry transformation.
Here, for simplicity, we neglect the gauge interactions
in the observable sector because such interactions are
not important when we consider the strong dynamics in the
hidden sector.
The Higgs fields $H_{i}$ does not have any $U(1)_{R}$ charge.
In general, the explicit mass parameter $m$ should be the order
of the Planck mass $M_{pl}$.
Other contributions to the superpotential, such as the non-renormalizable
couplings suppressed by the Planck mass, are not important in this model.
The classical equation of motion for the auxiliary component of $\Phi$
({$\phi,\chi,h$}) yields:
\begin{equation}
  \frac{\partial W}{\partial \phi}+\frac{1}{4}
  \frac{\partial f}{\partial \phi}\lambda\lambda=0
\end{equation}
This gives the relation:
\begin{equation}
  \label{eqnmo}
  (m^{2}+\lambda H_{1}H_{2})\phi=\frac{\xi}{4}\lambda\lambda.
\end{equation}
Once gaugino condensate in the hidden sector, the right hand side
of (\ref{eqnmo}) becomes about $\Lambda_{HS}^{3}$
where $\Lambda_{HS}$ is the dynamical scale of the hidden sector
($\Lambda_{HS}\sim 10^{12}$Gev).
For $H_{1}$ and $H_{2}$, it is obvious that their vacuum expectation
values cannot become large once $\phi$ develops non-zero vacuum
expectation value.
As a result, from the equation (\ref{eqnmo}) we obtain a solution
$\phi\sim <\lambda\lambda>/m^{2} \sim m_{\frac{3}{2}}$.
For the superfield $H_{i}$, this induces a supersymmetric mass term
of the electroweak scale.
\begin{equation}
  \label{mu}
  W^{\mu-term}\sim m_{\frac{3}{2}}H_{1}H_{2}
\end{equation}
This explains the scale of the $\mu$-term in a natural way.
The scalar component of $\Phi$ is determined mainly from the strong coupling
effect which is relevant to the gaugino condensation in the hidden sector,
and its non-zero value induces a supersymmetric mass term
to the Higgs superfields  which appears as
the $\mu$-term in the observable sector.
\section{Conclusion}
\hspace*{\parindent}
The $\mu$-problem is the necessity of introducing by hand
a small mass term($\mu$) of order the soft breaking
mass scale in the observable sector which (in general) is not
correlated to the breaking of supergravity.
We have shown that this problem can be avoided if we consider
an effective Lagrangian which possesses $U(1)_{R}$ symmetry
which is compensated by an auxiliary superfield $\Phi$.
\section*{Acknowledgment}
\hspace*{\parindent}
We thank K.Fujikawa and K.Tobe for many helpful discussions.


\begin{thebibliography}{1}
\bibitem{review}
J.E.Kim and H.P.Nills, Phys.Lett.B(1984)150\\
G.F.Giudice and A.Masiero, Phys.Lett.B(1988)480\\
I.Antoniadis, E.Gava, K.S.Narain and T.R.Taylor, Nucl.Phys.B432(1994)187\\
J.A.Casas and C.Mu\~unoz, Phys.Lett.B(1993)288\\
V.S.Kaplunovsky and J.Louis, Phys.Lett.B(1993)269\\
\bibitem{banks}
T.Banks, D.B.Kaplan and A.E.Nelson, Phys.Rev.D49(1994)779
\bibitem{ross}
A.de la Maccora and G.G.Ross, Nucl.Phys.B404(1993)321; \\
Phys.Lett.B325(1994)85; Nucl.Phys.B443(1995)127

\end{thebibliography}
\end{document}